\documentclass[preprint,showpacs,superscriptaddress,groupedaddress,nofootinbib]{revtex4-1}
\usepackage{graphicx}
\usepackage{bm}       
\usepackage{amssymb}
\usepackage[toc,page]{appendix}
\usepackage{braket}
\newcommand{\del}{\partial}

\usepackage{amsmath}

\begin{document}

\title{4D Einstein-Gauss-Bonnet Gravity From Non-Einsteinian Phase}
%4D Einstein-Gauss-Bonnet Gravity Generated By Invisible Extra Dimensions}

\author{Sandipan Sengupta}
\email{sandipan@phy.iitkgp.ac.in}
\affiliation{Department of Physics, Indian Institute of Technology Kharagpur, Kharagpur-721302, INDIA}

\begin{abstract}

We set up an Einstein-Gauss-Bonnet theory in four dimensions, based on the recent formulation of pure gravity with extra dimensions of vanishing metrical length \cite{sengupta}.
%We show that the Gauss-Bonnet and higher Lovelock terms become dynamically nontrivial in four dimensions within a recent formulation of gravity \cite{sengupta}, based on extra dimensions of vanishing metrical length. 
In absence of torsion, the effective field equations depend only on the four-metric, reflecting a quadratic curvature nonlinearity but no higher than second order derivatives.
In contrast with recent proposals to obtain 4dEGB theories through a singular rescaling of the Gauss-Bonnet coupling, this formalism requires no (classical) regularization of divergences and is inequivalent to Horndeski gravity, while being generally covariant and independent of compactification.
Notably, the vacuum field equations admit FLRW cosmologies containing nonsingular bounce and self-accelerating Universe, and spherically symmetric black holes more general than Schwarzschild. The quadratic theory of gravity emerging here is shown to be preserved against the inclusion of higher Lovelock densities in $5+2n$ dimensions.

%In presence of extra dimensions of vanishing proper length, we obtain a formulation of gravity theory where the Gauss-Bonnet
%density becomes dynamically nontrivial in four dimensions. 
%In absence of torsion, the effective field equations are given purely by the four metric with no higher than second order derivatives. The formalism is independent of compactification, requires no (classical) regularization of divergences and is generally covariant. This is in contrast with a recent proposal by Glavan and Lin as well as with prescriptions based on singular rescaling of couplings, both of which reproduce  scalar-tensor gravity in general. The vacuum field equations admit FLRW cosmologies containing nonsingular bounce and self-accelerating Universe, and spherically symmetric black holes superceding Schwarzschild. The effective theory, which reflects at the most a quadratic curvature nonlinearity, is preserved against the inclusion of higher order Lovelock densities. 

\end{abstract}

\maketitle
%%%%%%%%%%%%%%%%%%%%%%%%%%%%%%%%%%%%%%%%%%%%%%%%%%%%%%%%%%%%%%%%%%%%%%%

\section{Introduction}
The Gauss-Bonnet and higher Lovelock densities, as they are, do not affect the gravitational dynamics in four dimensions \cite{lovelock,lanc}. 
%To be precise, the (quadratic) Gauss-Bonnet density is topological and the higher order ones are trivial in four dimensions. 
This implies that there is no natural way to define a four dimensional gravity theory beyond Einstein's, where the field equations could reflect dynamical imprints of higher Lovelock terms through curvature nonlinearities without exhibiting additional degrees of freedom or higher than second derivatives of metric.

Apparently, the best one could do is to define a Kaluza-Klein reduction \cite{kaluza,klein} of a higher $(D>4)$ dimensional Einstein Gauss-Bonnet action \cite{hoissen,deruelle,felisola}. 
%The resulting four dimensional effective theory is very much dependent on the choice of compactification . 
In general, this leads to scalar-vector-tensor theories of gravity where the non-metric propagating modes encode the dynamical effect of extra dimensions. 
Recently though, it had been claimed that by introducing a singular rescaling of the Gauss-Bonnet coupling and then imposing a $D\rightarrow 4$ limit, one could reinterpret the original $D\geq 5$ dimensional solutions of Einstein-Gauss-Bonnet action \cite{wheeler} as four dimensional `solutions'  of a new metric theory beyond Einstein's \cite{glavan}. 
%Apparently, these formulation reflects the dynamical imprint of a Gauss-Bonnet correction without introducing any new degree of freedom }. 
However, such a formulation admits neither an action formulation nor a set of field equations that are well-defined in the four dimensional limit. To elaborate, the formalism is not generally covariant, not unique (the emergent theory depends on the dynamics of extra dimensions even though these are treated as fiducial in the limit) and has a number of  conceptual drawbacks essentially stemming from the ill-defined limit \cite{tekin,arrechea,aoki}. In retrospect, a complete consistency of such a proposal inspite of the manifest violation of Lovelock's theorem would imply that there should be no essential distinction between dynamical and topological densities in a given number of dimensions. Such a feature would be fairly disturbing. 

In fact, the configurations discussed by Glavan and Lin have subsequently been observed to be solutions to the Horndeski class \cite{horndeski} in four dimensions, which does exhibit an additional propagating scalar contrary to the original claim \cite{lu}. 
These theories could be derived using a Kaluza-Klein reduction on a maximally symmetric space whose breathing mode precisely corresponds to this scalar. However, this compactification procedure leads to effective four-dimensional theories that are not unique in general.

As an alternative method to reproduce these scalar-tensor gravity theories, a conformal scaling prescription has also been invoked \cite{mann,clifton}. However, such proposals are defined upon a singular rescaling of the Gauss-Bonnet coupling in a bimetric action, along with the subtraction of an infinite total derivative. The physical interpretation of this purely classical regularization prescription remains unclear. 
%the conformal trick, while being independent of compactification, requires two different metrics along with the corresponding dynamical terms in a (bimetric) action to begin with and ends up with a single one after the regularized limit is imposed. Moreover, 
Moreover, the issue whether the discontinuous limit $D\rightarrow 4$ applied to $(D-4)$-dimensional tensors could make sense without leading to ambiguous final results, as the extra dimensions are eventually forced to be treated as fictitious, is a rather subtle one. 
In fact, the solutions for a D-dimensional gravity theory with an $n<D$ dimensional subspace of vanishing metrical size (corresponding to a $D$-metric having $n$ zero eigenvalues) is not equivalent to one where these same $n$ directions define an infinitesimally small proper volume (in the limit $D\rightarrow 4$). For instance, in the context of first order gravity in four dimensions ($D=4$), it has already been noted earlier that these two sectors are related by singular diffeomorphisms \cite{horowitz}, and lead to quite different solution spaces \cite{tseytlin,kaul}.
%In general, all possible higher Lovelock terms should contribute to these effective theories in an arbitrary number of dimensions, leading to a highly nonlinear structure unless some additonal physics is invoked to truncate it.

Thus, a new framework to find an effective four dimensional theory is necessary, if the formalism has to reflect genuine dynamical effects induced by higher Lovelock terms without depending upon the dynamics associated with extra dimensions, or requiring any singular rescaling of couplings followed by a classical regularization of divergent actions. If such a formulation could reflect a purely gravitational sector associated with field equations having no higher than second derivatives (of metric), that should indeed signify some progress.

Here we elucidate such a framework. Its basis is a non-Einsteinian phase of higher dimensional gravity in vacuum, introduced recently \cite{sengupta}. In contrast with the Kaluza-Klein proposal \cite{kaluza,klein}, the fifth dimension in this formalism represents a vanishing eigenvalue of the (non-invertible) five-vielbein. Having a trivial metrical length, this extra dimension has no dynamics of its own, and cannot be detected in principle.
In ref.\cite{sengupta}, this formulation was applied to set up an effective four dimensional theory derived from the 
five dimensional Hilbert-Palatini action. Based on the non-particulate nature of the emergent (torsional) non-propagating fields which exhibit no inherent mass scale, these were  proposed to supercede the hypothetical `dark matter'. It was also shown that any possible coupling of these fields with genuine matter fields in the four dimensional theory are necessarily Planck suppressed, and their characteristic length scales precisely correspond to the scale of Galactic accelerations which in turn is related to the cosmological constant. These are quite distinctive features, defining any potential `dark matter' candidate. This proposal predicts that the equation of state of the `dark' geometric multiplet must be bounded as $-\frac{1}{3}\leq w \leq 1$, where the lower limit is precisely what is required to explain the non-Newtonian features of the galactic halo corresponding to flat rotation curves.
 
%The resulting  effective theory was aimed towards addressing the `dark matter' problem. 
Here, we consider a dynamical formulation based on a five dimensional action with Einstein and Gauss-Bonnet terms, where the null eigenvalue of the five-vielbein (metric) represents the `dark' extra dimension. The resulting four dimensional effective field equations are shown to inherit nonlinear curvature effects induced by the Gauss-Bonnet density, thus providing an intriguing contrast with four dimensional pure Einstein gravity. 

A generalization to the case with more than one dimension of zero proper length is also presented. This reveals the remarkable fact that Lovelock terms higher than Gauss-Bonnnet do not affect the four dimensional gravity theory obtained earlier with only one vanishing metrical dimension. 

The question of a regularization in a discontinuous limit $D\rightarrow 4$ does not arise at all in this formulation.  The resulting field equations depend only on the four-metric and its derivatives upto second order through the (torsionfree) curvature tensor. This is in contrast with the Kaluza-Klein reduced theories obtained from a higher dimensional Einstein-Gauss-Bonnet (Lovelock) action as well as to the ones obtained through the singular $D\rightarrow 4$ prescription, where the additional scalar degrees of freedom do not decouple from the
emergent metric \cite{lu,clifton}. 

The relevance of our formalism could go beyond the motive of constructing an effective Einstein-Gauss-Bonnet metric theory. We note that the only nonlinear correction induced by the full Lovelock series in the effective four dimensional theory is quadratic, and is unique in form. The inherent economy in the emergent structure is expected to reduce the ambiguity pervading higher curvature gravity theories in general{\footnote{Even though physically reasonable requirements such as ghost freedom etc. have been considered earlier \cite{biswas}, these still allow for an infinity of terms in the equations of motion, in contrast with our formalism.}}. Among the non-Einsteinian signatures seeded by the nonlinearities in this vacuum theory, the emergence of  nonsingular bouncing cosmology \cite{steinhardt,biswas1} even in absence of a spatial curvature and a bare cosmological constant is intriguing. In the case of (static) spherical symmetry, the Schwarzschild solution of Einstein gravity is superceded by a four-parameter vacuum solution consisting of an emergent mass and charge, exhibiting curvature singularity. 

In the next section, we begin with a discussion of the basic five dimensional action and the resulting field equations. This is followed by analyses of the emergent four dimensional theory resulting from their solutions, in particular given by spacetimes with constant curvature and vanishing torsion, respectively. The dynamical implications for FLRW cosmology and static spherically symmetric spacetimes are discussed. The generalization to higher order Lovelock densities in presence of additional dimensions of zero proper length is also presented. The final section contains a summary and criticism of our work. 
%Among the notable features, we find that the nonlinearities could seed nonsingular bouncing solutions even in the absence of a cosmological constant. 
%In the case of static spherically symmetric four geometries, the Schwarzschild solution of vacuum Einstein gravity is superceded by a more general one which, nevertheless, exhibits curvature singularities. These exhibit an emergent Reissner-Nordstrom charge, in contrast to the solutions obtained earlier from Kaluza-Klein dimensional reductions of Einstein-Gauss Bonnet or through a four dimensional limit (upto regularization) of higher dimensional solutions \cite{wheeler,lu}.

%%%%%
%%%%%%
\section{The fundamental theory}

The fundamental five-dimensional Lagrangian density is given by:
\begin{eqnarray}\label{L}
{\cal L}(\hat{e},\hat{w})&=&\epsilon^{\mu\nu\alpha\beta\gamma} \epsilon_{IJKLM}\left[\frac{\alpha}{2}~\hat{R}_{\mu\nu}^{~~IJ}(\hat{w})\hat{R}_{\alpha\beta}^{~~KL}(\hat{w})\hat{e}_{\gamma}^{M}~+~\frac{\zeta}{3}~\hat{R}_{\mu\nu}^{~~IJ}(\hat{w})\hat{e}_{\alpha}^{K}\hat{e}_{\beta}^{L}\hat{e}_{\gamma}^{M} ~+~\frac{\beta}{5}~\hat{e}_{\mu}^{I}\hat{e}_{\nu}^{J}\hat{e}_{\alpha}^{K}\hat{e}_{\beta}^{L}\hat{e}_{\gamma}^{M}\right],\nonumber\\
&&
\end{eqnarray}
where the vielbein $\hat{e}_\mu^I(x)$ and the super-connection $\hat{w}_\mu^{~IJ}(x)$ ($\mu\equiv [t,x,y,z,v]$, $I\equiv [0,1,2,3,4]$) are the independent fields. $\zeta$ and $\alpha$ define the gravitational and Gauss-Bonnet coupling, respectively and $\beta$ is the bare cosmological constant. These have dimensions: $\zeta\sim M^{-1}L^2,~\alpha\sim M,~\beta\sim ML^{-4}$. The $SO(4,1)$ field-strength is defined as $\hat{R}_{\beta\rho}^{~~LM}(\hat{w})=\del_{[\beta} \hat{w}_{\rho]}^{~LM}+\hat{w}_{[\beta}^{~LK}\hat{w}_{\rho]K}^{~~~M}$, the internal metric being defined as $\eta_{IJ}=[-1,1,1,1,1]$. 

A variation of (\ref{L}) with respect to the connection and the vielbein leads to the following field equations in vacuum, respectively: 
\begin{subequations}
\begin{align}
\epsilon^{\mu\nu\alpha\beta\gamma} \epsilon_{IJKLM}\left[\zeta\hat{e}_{\mu}^{I} \hat{e}_{\nu}^{J} +\alpha\hat{R}_{\mu\nu}^{~IJ}(\hat{w})\right] \hat{D}_{\alpha}(\hat{w})\hat{e}_{\beta}^{K}=0,\label{eom1}\\
\epsilon^{\mu\nu\alpha\beta\gamma} \epsilon_{IJKLM}\left[\zeta\hat{e}_{\mu}^{I} \hat{e}_{\nu}^{J} \hat{R}_{\alpha\beta}^{~KL}(\hat{w})+\frac{\alpha}{2}\hat{R}_{\mu\nu}^{~IJ}(\hat{w})\hat{R}_{\alpha\beta}^{~KL}(\hat{w})+\beta\hat{e}_{\mu}^{I} \hat{e}_{\nu}^{J} \hat{e}_{\alpha}^{K} \hat{e}_{\beta}^{L} \right]=0\label{eom2}
\end{align}
\end{subequations}
Here we have defined $\hat{D}_\mu(\hat{w})$ as the gauge-covariant derivative with respect to the super-connection $\hat{w}_\mu^{~IJ}$ and have used the Bianchi identity $\hat{D}_{[\mu}\hat{R}_{\nu\alpha]}^{~~IJ}(\hat{w})=0$ in obtaining the connection equations (\ref{eom1}).

In general, both the Lagrangian density (\ref{L}) as well as these equations above are well-defined for invertible as well as non-invertible vielbein. The invertible spacetime solutions, upon a compactification of the fifth dimension, is equivalent to the well-known Kaluza-Klein reduction of the five dimensional Einstein Gauss-Bonnet action. Here, however, we shall be concerned with solutions with a degenerate vielbein (one zero eigenvalue), leading to a solution space completely inequivalent to the invertible five-metric case.
%\footnote{The mutual inequivalence of the invertible and non-invertible spacetime solutions in first order gravity in four dimensions have been noted and explored in references \cite{tseytlin,kaul}.}.  
As is obvious from the analysis below, one does not require any additional stabilization mechanism for the extra dimension, which itself cannot be detected directly as it is not associated with any genuine dynamics. 

The connection equations (\ref{eom1}) are solved by:\\
 a) $\zeta\hat{e}_{\mu}^{I} \hat{e}_{\nu}^{J} +\alpha\hat{R}_{\mu\nu}^{~IJ}(\hat{w})=0$;\\ 
 b) $\hat{D}_{\alpha}(\hat{w})\hat{e}_{\beta}^{K}=0$.\\
In what follows next, we shall explore these cases and present the complete solutions leading to the four dimensional gravity theory, before elucidating specific applications.

We adopt a notation similar to ref.\cite{sengupta}, where the notion of zero length extra dimensions had been introduced. The zero eigenvalue of the vielbein $\hat{e}_\mu^I$ could be chosen to  lie along the fifth dimension ($v$):
\begin{eqnarray*}
\hat{e}_v^I=0,
\end{eqnarray*}
leading to:
\begin{eqnarray}\label{e}
\hat{e}_\mu^I =
\left[\begin{array}{cc}
\hat{e}_a^i\equiv e_a^i & 0 \\
0 & 0 \\
\end{array}\right]
\end{eqnarray}
Here the world indices are $\mu\equiv (a,v)\equiv (t,x,y,z,v)$ and the internal indices are $I\equiv (i,4)=(0,1,2,3,4)$.

The emergent tetrad fields $e_a^i$ (invertible) define the effective four-dimensional spacetime.  Their inverse are denoted as $e^a_i$ (not the same as $\hat{e}^a_i$ which do not exist):
\begin{eqnarray*}
e_a^i e^b_i=\delta_a^b,~e_a^i e^a_j=\delta _j^i.
\end{eqnarray*}
In terms of these, the corresponding $4$-metric is defined as $g_{ab}=e_a^i e_{bi}=\hat{g}_{ab}$ which are the only nontrivial components of the fundamental five-metric $(\hat{g}_{va}=\hat{g}_{av}=0=\hat{g}_{vv})$. The four dimensional epsilon symbols are derived from the five dimensional antisymmetric tensor densities as: $\epsilon^{vabcd}\equiv \epsilon^{abcd},~\epsilon_{4ijkl}\equiv\epsilon_{ijkl}$.

\section{Constant curvature spacetime solutions}
Here we consider the class (a) of spacetime solutions:
\begin{eqnarray}\label{constk}
\zeta\hat{e}_{\mu}^{I} \hat{e}_{\nu}^{J} +\alpha\hat{R}_{\mu\nu}^{~IJ}(\hat{w})=0
\end{eqnarray}
The requirement that such spacetimes solve the vielbein equations of motion (\ref{eom2}) implies the following constraint among couplings:
\begin{eqnarray}\label{coupling}
\zeta^2-2\alpha\beta=0.
\end{eqnarray}
Using the non-invertible vielbein (\ref{e}), decomposition of the five dimensional equations (\ref{constk})  leads to:
\begin{eqnarray}\label{R}
\hat{R}_{av}^{~i4}&=&\del_{[a}\hat{w}_{v]}^{~i4}+\hat{w}_{[a}^{~ik}\hat{w}_{v]k}^{~~~4}=0,\nonumber\\
\hat{R}_{av}^{~ij}&=&\del_{[a}\hat{w}_{v]}^{~ij}+\hat{w}_{[a}^{~ik}\hat{w}_{v]k}^{~~~j}+\hat{w}_{[a}^{~i4}\hat{w}_{v]4}^{~~~j}=0\nonumber\\
\hat{R}_{ab}^{~i4}&=&\del_{[a}\hat{w}_{b]}^{~i4}+\hat{w}_{[a}^{~ik}\hat{w}_{b]k}^{~~~4}=0\nonumber\\
\hat{R}_{ab}^{~ij}&+&\frac{\zeta}{2\alpha}e_{[a}^i e_{b]}^j=0=\bar{R}_{ab}^{~ij}(\bar{w})+\bar{D}_{[a}(\bar{w})K_{b]}^{~ij}+K_{[a}^{~ik} K_{b]k}^{~~j}+\hat{w}_{[a}^{~i4}\hat{w}_{b]4}^{~~~j}+\frac{\zeta}{2\alpha} e_{[a}^i e_{b]}^j
\end{eqnarray}
In the last equation above we have used the fact that the connection components $\hat{w}_a^{~ij}$ could in general be decomposed as:
\begin{eqnarray*}
\hat{w}_a^{~ij}=\bar{w}_a^{~ij}(e)+K_a^{~ij},
\end{eqnarray*}
where $\bar{w}_a^{~ij}(e)=\frac{1}{2}[e^b_i\del_{[a}e_{b]}^j
-e^b_j\del_{[a}e_{b]}^i -  e_a^l e^b_i e^c_j
\del_{[b}e_{c]}^l]$ defines the torsionless connection ($\bar{D}_{[a}(\bar{w})e_{b]}^i=0$) and  $K_a^{~ij}=-K_a^{~ji}$ is the contortion tensor in the emergent spacetime. To emphasize, the covariant derivative $\bar{D}_{a}(\bar{w})$ is defined with respect to the connection $\bar{w}_a^{~ij}$.

\subsection*{Solution for connection fields:}

In order to unravel the full implications of the above equations of motion, we first analyze the Bianchi identities: $\epsilon^{\mu\nu\alpha\beta\gamma} \epsilon_{IJKLM}\hat{e}_{\mu}^{I}\hat{D}_{\nu} \hat{R}_{\alpha\beta}^{~JK}=0$. For the spacetime solutions (\ref{constk}), these reduce to:
\begin{eqnarray}\label{bian1}
\epsilon^{\mu\nu\alpha\beta\gamma} \epsilon_{IJKLM}\hat{e}_{\mu}^{I}\hat{e}_\nu^J \hat{D}_{\alpha}\hat{e}_\beta^K =0
\end{eqnarray}
These are linear in the connection fields, whose solutions are found below.

From the $\gamma=v,L=l,M=m$ component of eq.(\ref{bian1}), we obtain:
\begin{eqnarray}
\epsilon^{abcd} \epsilon_{jklm}e_{a}^{j}e_{b}^{k} \hat{D}_{c}e_{d}^{4}=0=\hat{w}_{[a}^{~4j}e_{b]j}
\Rightarrow \hat{w}_{a}^{~4j}=e_{ai}M^{ij},
\end{eqnarray}
where $M^{ij}=M^{ji}$ is a $4\times 4$ matrix field, arbitrary upto the equations of motion.
The component $\gamma=v,L=l,M=4$ implies:
\begin{eqnarray}
\epsilon^{abcd} \epsilon_{ijkl}e_{a}^{i}e_{b}^{j} \hat{D}_{c}e_{d}^{k}=0=e^{c}_k \hat{D}_{[b}e_{c]}^k
\Rightarrow e^a_k K_a^{~ik}=0=K_{ai}^{~a},
\end{eqnarray}
which forces four of the twenty four contortion components to vanish.
Next, the component $\gamma=d,L=l,M=m$ reads:
\begin{eqnarray}
\epsilon^{abcd} \epsilon_{ijlm}e_{a}^{i}e_{b}^{j} \hat{D}_{v}e_{c}^{4}=0=e^{c}_{[k} e^{d}_{l]}\hat{w}_{v}^{~4m}e_{c]m}
\Rightarrow \hat{w}_{v}^{~4m}=0.
\end{eqnarray}
Finally, for $\gamma=d,L=l,M=4$ we find:
\begin{eqnarray}
\epsilon^{abcd} \epsilon_{ijkl}e_{a}^{i}e_{b}^{j} \hat{D}_{v}e_{c}^{k}=0=e^{c}_{[k} e^{d}_{l]}\hat{D}_{v}e_{c}^k
\Rightarrow e^c_k \hat{D}_v e_c^k=0=\hat{D}_v e_a^i.
\end{eqnarray}
From this, it follows that the determinant of the tetrad is $v$-independent: $\del_v e=0$ and that $\hat{w}_v^{~ij}$ is a pure gauge:
\begin{eqnarray*}
\hat{w}_v^{~ij}=-e^a_j\del_v e_a^i
\end{eqnarray*} 
As a consequence, the emergent 4D metric $g_{ab}$ exhibits no dependence on the fifth coordinate $v$:
$\del_{v}g_{ab}=\del_v (e_a^i e_{bi})=e_{ai} \hat{D}_v e_b^i+(a\leftrightarrow b)=0$.
All these facts together imply that any possible dependence of the tetrad on $v$ must be a gauge artifact and might just be gauged away by choosing the trivial gauge:
\begin{eqnarray}
\hat{w}_v^{~ij}=0.
\end{eqnarray}

\subsection*{Four dimensional effective theory:}

Inserting these solutions for the connection fields found above into the first three equations in the set (\ref{R}), we obtain the following constraints:
\begin{eqnarray}\label{R1}
\hat{R}_{av}^{~i4}&=&0=\del_v M^{ik},~\hat{R}_{av}^{~ij}=0=\del_v K_a^{~ij},\nonumber\\
\hat{R}_{ab}^{~i4}&=&0=\bar{D}_{[a}(\bar{w})M_{b]}^i+K_{[a}^{~ij}M_{b]j}
\end{eqnarray}
where we have defined: $M_a^i=M^{ij}e_{aj}$.
As expected from the arguments presented in the previous  paragraph, the fields $M^{ij},K^{~ij}_a$, which are the only emergent fields other than the tetrad, also exhibit no dependence on the fifth coordinate associated with the null eigenvalue. The last equation, however, is intriguing in its own right and could be interpreted further. Since the tensor $M_{ij}$ is symmetric, this equation could be solved for the contortion in terms of $M_{ij}$ itself along with its inverse and first derivative. Note that the field $M_{ab}=M_{ij}e_a^i e_b^j$ could then be seen as a second emergent metric. From the last equation in (\ref{R}), it is evident that both metrics would be dynamical in general. Hence, this formulation admits an interpretation as an emergent bimetric theory of gravity. 
However, here we shall not explore this interesting direction any further, as it is beyond the scope and purpose of this article. Rather, we assume that there is only a single emergent metric and hence the tensors $g_{ab}$ and $M_{ab}$ must be proportional:
\begin{eqnarray}\label{M}
M_a^i=\lambda e_a^i, ~\lambda\equiv const.
\end{eqnarray} 
This, when inserted back into eq.(\ref{R1}), implies:
\begin{eqnarray}
K_{[a}^{~ij}e_{b]j}=0\Rightarrow K_a^{~ij}=0.
\end{eqnarray}
In other words, the four dimensional torsion must vanish. 
%It could be of interest to contrast this case to the one without the Gauss-Bonnet term \cite{sengupta}, where torsion could be nonvanishing for the solution (\ref{M}).

Using the results above, the only remaining equation of motion, given by the last one in the set (\ref{R}), finally simplifies to:
\begin{eqnarray}
\bar{R}_{ab}^{~ij}(\bar{w}(e))+\left[\frac{\zeta}{2\alpha} -\lambda^2\right]e_{[a}^i e_{b]}^j=0.
\end{eqnarray} 
These solutions represent maximally symmetric four dimensional spacetimes.
The curvature is positive, negative or zero provided: $\frac{\zeta}{2\alpha}<\lambda^2$, $\frac{\zeta}{2\alpha}>\lambda^2$ or $\frac{\zeta}{2\alpha}=\lambda^2$, respectively. The only vestige of the Gauss-Bonnet contribution is contained in this spacetime curvature through its coupling $\alpha$.

This completely defines the four dimensional gravity theory resulting from the solution (\ref{constk}), described by a single emergent metric and no other propagating fields.
%The solutions obtained above completely specify the four dimensional spacetime geometry corresponding to the case with a single emergent tetrad field ($e_a^i$) and no other propagating degrees of freedom. 

In the next section, we explore whether the general solution leads to a gravitational dynamics exhibiting nonlinear curvature effects.

\section{Spacetime solutions with vanishing torsion}

As discussed earlier, the connection equations exhibit yet another important class of solutions, which exhibits lesser symmetry but underlies a richer structure than the previous case:
\begin{eqnarray}\label{5dT}
\hat{D}_{[\alpha}(\hat{w})e_{\beta]}^{I}=0.
\end{eqnarray}
%This class is what shall concern us in the rest of this work.
Let us proceed to find the most general solutions following from the above.

Rewriting eq.(\ref{5dT}) in terms of the four dimensional emergent fields, we obtain the following solutions for the five dimensional spin-connection components:
\begin{eqnarray}
\hat{D}_{[a}(\hat{w})e_{b]}^{i}&=&0\Rightarrow K_a^{~ij}\equiv \hat{w}_a^{~ij}-\bar{w}_a^{~ij}(e)=0,\nonumber\\
\hat{D}_{[a}(\hat{w})e_{b]}^{4}&=&0\Rightarrow \hat{w}_a^{~4i}=e_{ak} Q^{ik}\equiv Q_a^i ~[Q^{ik}=Q^{ki},]\nonumber\\
\hat{D}_{[v}(\hat{w})e_{a]}^{4}&=&0\Rightarrow \hat{w}_v^{~4i}=0,\nonumber\\
\hat{D}_{[v}(\hat{w})e_{a]}^{i}&=&0\Rightarrow \hat{w}_v^{~ij}=-e^a_j\del_v e_a^i.
\end{eqnarray}
Exactly as in the earlier section, the last equation implies that $\hat{w}_v^{~ij}$ is a pure gauge. Once again, we choose the physical gauge $\hat{w}_v^{~ij}=0$ which eliminates the spurious $v$-dependence of the tetrad. The only nontrivial connection component is encoded by the symmetric emergent field $Q_{kl}$ (or equivalently, $Q_a^i$).

Let us analyze the vielbein equations of motion (\ref{eom2}) next. Note that the components $(\gamma=a,M=4)$, $(\gamma=a,M=i)$ are identically satisfied upon using the identities $\hat{R}_{va}^{~4j}=0=\hat{R}_{va}^{~ij}$ which follow from the solutions obtained earlier. The $\gamma=v,M=i$ component reads:
\begin{eqnarray}\label{RR1}
\epsilon^{abcd} \epsilon_{ijkl}\left[\hat{R}_{ab}^{~kl}+\frac{\zeta}{\alpha}e_{a}^{k}e_{b}^{l}\right]\hat{R}_{cd}^{~4j}=0=\epsilon^{abcd} \epsilon_{ijkl}\left[\bar{R}_{ab}^{~kl}-2Q_{a}^{k}Q_{b}^{l}+\frac{\zeta}{\alpha}e_{a}^{k}e_{b}^{l}\right]\left(\bar{D}_c(\bar{w})Q_d^j\right)
\end{eqnarray}
Evidently, this is a cubic equation in $Q_a^i$. As in the previous section, the emergent symmetric tensor $Q_{ab}=Q^{ij}e_{ai}e_{bj}$ admits an interpretation as a dynamical metric other than $g_{ab}$. Such a possibility would not be explored any further here though. 

Let us consider the case where there is only one emergent metric. The solution to eq.(\ref{RR1}) satisfying this property is given by:
\begin{eqnarray}\label{Q}
Q_a^i=\lambda e_a^i, ~\lambda\equiv const.~~,
\end{eqnarray}
noting that this satisfies the identity $\bar{D}_{[c}(\bar{w})Q_{b]}^i=0$. 
%Note that the other particular solution simply reduces to the case of constant curvature spacetimes already analyzed in detail earlier.

The only remaining component of the field equations (\ref{eom2}) is $\gamma=v,M=4$, resulting in the following scalar constraint:
 \begin{eqnarray}\label{eom}
\epsilon^{abcd} \epsilon_{ijkl}\left[\zeta\hat{R}_{ab}^{~ij}e_{c}^{k}e_{d}^{l}+\frac{\alpha}{2}\hat{R}_{ab}^{~ij}\hat{R}_{cd}^{~kl}+\beta 
e_{a}^{i}e_{b}^{j}e_{c}^{k}e_{d}^{l}\right]= 0=\epsilon^{abcd} \epsilon_{ijkl}\left[\phi\bar{R}_{ab}^{~ij}e_{c}^{k}e_{d}^{l}+\frac{\alpha}{2}\bar{R}_{ab}^{~ij}\bar{R}_{cd}^{~kl}+\chi e_{a}^{i}e_{b}^{j}e_{c}^{k}e_{d}^{l}\right],\nonumber\\
~
\end{eqnarray}
where in the last equality we have redefined the coupling constants $(\zeta,\alpha,\beta)$ as $(\phi,\alpha,\chi)$ with $\phi=(\zeta-2\alpha\lambda^2) $ and $\chi=(\beta-2\zeta\lambda^2+2\alpha \lambda^4)$. The simple equation above, along with the Bianchi identities, constitute the full system of equations of motion and defines the effective Einstein-Gauss-Bonnet theory. The analysis of Bianchi identities, which eventually demonstrates the $v$-independence of the nontrivial emergent connection components, is very similar to the one presented at the earlier section and would not be repeated here.

Note that the maximally symmetric spacetimes of section-III are particular solutions to the quadratic gravity theory defined by the field equations(\ref{eom}). Hence, the (more general) class of vanishing torsion solutions is what shall concern us in the rest of this work.

Some of the notable features reflected by the four dimensional theory obtained above are outlined below:

(a) The emergent equation of motion (\ref{eom}) is inequivalent to Einstein's equations. While it inherits curvature nonlinearities induced by the Gauss-Bonnet term, it has no higher than second derivatives of the four-metric, and hence is free of Ostrogradsky instability.

(b) The field equations depend only on the metric and its curvature. The resulting theory is strictly inequivalent to  scalar-tensor gravity, as obtained from either a Kaluza-Klein reduction or a `dimensional regularization' of a higher dimensional Gauss-Bonnet theory  \cite{hoissen,glavan,lu,clifton}. To emphasize, in these Kaluza-Klein reduced effective theories, the propagating scalar mode does not decouple from gravitational sector, hence leading to the trivial case (Einstein gravity) when taken away.

(c) The Bianchi identity $\epsilon^{abcd}\hat{D}_b (\hat{w})R_{cd}^{~ij}(\hat{w})=0 \implies\nabla_a\left[\bar{R}^{ab}(\bar{w}(e))-\frac{1}{2}g^{ab}\bar{R}(\bar{w}(e))\right]=0$ reflects the general covariance of the emergent gravity theory ($\nabla_a$ being the covariant derivative defined with respect to the Christoffel symbol).

(d) The equations of motion have been obtained without using any singular rescaling of the coupling constants or any regularization involving a discontinuous limit such as  $D\rightarrow 4$. This may be contrasted with the recent proposal by Glavan and Lin \cite{glavan} and a series of subsequent ones based on a classical regularization of a divergent action defined by two conformally related metrics \cite{clifton}.

(e) The equation of motion (\ref{eom}) could be derived from a well-defined four dimensional action principle involving an auxiliary scalar field $\psi$:
\begin{eqnarray*}
{\cal L}_4(e,\omega,\psi,\Lambda)=\psi\epsilon^{abcd} \epsilon_{ijkl}\left[\phi R_{ab}^{~ij}(\omega)e_{c}^{k}e_{d}^{l}+\frac{\alpha}{2}R_{ab}^{~ij}(\omega)R_{cd}^{~kl}(\omega)+\chi e_{a}^{i}e_{b}^{j}e_{c}^{k}e_{d}^{l}\right]+\Lambda^{ab}_{~~i}D_{[a}(\omega)e_{b]}^i,
\end{eqnarray*}
where $e_a^i$ and $\omega_a^{~ij}$ are the tetrad and connection fields respectively of the purely four dimensional theory and $(\phi,\alpha,\chi)$ are coupling constants.
It is straightforward to verify that the multiplier fields $\psi$ and $\Lambda^{ab}_{~~i}=-\Lambda^{ba}_{~~i}$ decouple from the equation governing the metric dynamics, which is precisely equivalent to eq.(\ref{eom}) above.
 
%Thus, the emergent theory is indeed geometrical.\\
%(f) The formalism here is inequivalent to a Kaluza-Klein reduction of a higher dimensional Gauss-Bonnet theory or to its dimensionally regularized version discussed in the literature \cite{hoissen,glavan,lu,clifton}. Both of these latter classes are known to contain at least one additional propagating scalar mode which trivially leads to Einstein gravity when taken away.\\
(f) The fact that the induced four dimensional physics should not depend on the fifth coordinate, which is associated with a zero proper length $(\hat{g}_{v\mu}=0)$ and hence underlies no real dynamics, naturally follows from the original field equations. This may be contrasted with the Kaluza-Klein compactification where the requirement of the independence of the fifth coordinate needs to be imposed.

\section{FLRW cosmology}

As a first application of the formalism presented earlier, we shall elucidate the dynamical consequences of the effective field equations (\ref{eom}) in the case of homogeneous and isotropic cosmology. In particular, it is important to isolate the new features, if any, induced by nonlinearities in curvature which are absent in Einstein gravity. 

We idealize our considerations by assuming a homogeneous and isotropic form for the emergent spacetime ($k=0,\pm 1$ being the spatial curvature):
\begin{eqnarray*}
ds^2=-dt^2+a^2(t)\left[\frac{dr^2}{1-kr^2}+d\theta^2+\sin^2\theta d\phi^2\right],
\end{eqnarray*}
Using the identities involving the torsionfree curvature two-form:
\begin{eqnarray*}
\epsilon^{abcd}\epsilon_{ijkl}e_a^i e_b^j\bar{R}_{cd}^{~kl}(e)&=&\frac{24\left[a\ddot{a}+\dot{a}^2+k\right]ar^2\sin\theta}{\sqrt{1-kr^2},},\nonumber\\
\epsilon^{abcd}\epsilon_{ijkl}\bar{R}_{ab}^{~ij}\bar{R}_{cd}^{~kl}(e)&=&\frac{96\left[\dot{a}^2+k\right]\ddot{a}r^2\sin\theta}{\sqrt{1-kr^2}},
\end{eqnarray*} 
the equation of motion (\ref{eom}) becomes:
\begin{eqnarray}\label{frw1}
\phi\left[\frac{\ddot{a}}{a}+\frac{\dot{a}^2+k}{a^2}\right]+2\alpha\left[\frac{\dot{a}^2+k}{a^2}\right]\frac{\ddot{a}}{a}+\chi=0,
\end{eqnarray}

Let us first consider the spatially flat case ($k=0$). Introducing the variable $u=\frac{\dot{a}}{a}$, we rewrite (\ref{frw1}) as a first order equation:
\begin{eqnarray}
\phi\left[\dot{u}+2u^2\right]+2\alpha\left[\dot{u}+u^2\right]u^2+\chi=0.
\end{eqnarray}
Its general solution is given by:
\begin{eqnarray}
-t+const.=\sqrt{\frac{\alpha}{2}}\left[\frac{\tan^{-1}\left(\frac{\sqrt{2\alpha}u}{\sqrt{\phi-\sqrt{\phi^2-2\alpha\chi}}}\right)}{\sqrt{\phi-\sqrt{\phi^2-2\alpha\chi}}}+\frac{\tan^{-1}\left(\frac{\sqrt{2\alpha}u}{\sqrt{\phi+\sqrt{\phi^2-2\alpha\chi}}}\right)}{\sqrt{\phi+\sqrt{\phi^2-2\alpha\chi}}}\right]
\end{eqnarray}
This is a transcental equation, having no closed form solution for $a(t)$. However, the approximate solutions in the limit of small and large (with respect to the smallest and largest frequency scales, respectively, among $\sqrt{\frac{\phi}{2\alpha}\pm\sqrt{\frac{\phi^2}{4\alpha^2}-\frac{\chi}{2\alpha}}}$ ) Hubble rate $u$ are displayed below:
\begin{eqnarray*}
\mathrm{Small ~u~limit:}~~ &&a(t)\approx A e^{\frac{\chi}{2\phi}(Bt-t^2)}\nonumber\\
\mathrm{Large ~u~limit:}~~ &&a(t)\approx C (t-D)
\end{eqnarray*}
where $(A,B)$ and $(C,D)$ are the integration constants. The small $u$ limit could be expected to describe the late time dynamics, exhibiting what might be called a generalized de-Sitter (anti de-Sitter) behaviour with a nonlinear exponent. The large $u$ limit, on the other hand, suggests a spatially flat Milne cosmology at early times. Neither of these behaviours have an analogue in the Einsteinian case in general. Note that the theory is well-defined for $\frac{\phi^2}{2\alpha}-\chi\geq 0$.

Next, we summarize the main cosmological models emerging from the equation of motion (\ref{frw1}).

\subsection*{Non-oscillatory solutions:}

(a) $k=\pm 1$: $a(t)=Ae^{\mu t}+\frac{k}{4A\mu^2} e^{-\mu t}$, $\mu^2=-\frac{\phi}{2\alpha}\pm \sqrt{\left(\frac{\phi}{2\alpha}\right)^2-\frac{\chi}{2\alpha}}$.

(b) $k=+1$: $a(t)=\frac{1}{\mu}\cosh\mu t$, $\mu^2=-\frac{\phi}{2\alpha}\pm \sqrt{\left(\frac{\phi}{2\alpha}\right)^2-\frac{\chi}{2\alpha}}$. This represents a symmetric nonsingular bounce.

(c) $k=-1$: $a(t)=\frac{1}{\mu}\sinh\mu t$, $\mu^2=-\frac{\phi}{2\alpha}\pm \sqrt{\left(\frac{\phi}{2\alpha}\right)^2-\frac{\chi}{2\alpha}}$. 

(d) $k=-1$: $a(t)=t$, $\chi=0$. This is the Milne dynamics.

(e) $k=0$: $a(t)=A e^{\mu t}$, $\mu^2=-\frac{\phi}{2\alpha}\pm \sqrt{\left(\frac{\phi}{2\alpha}\right)^2-\frac{\chi}{2\alpha}}$ ($\phi\neq -2\alpha \mu^2$). This reflects a pure de-Sitter (anti de-Sitter) behaviour.

(f) $k=0$: $a(t)=Ae^{\mu t}+B e^{-\mu t}$, $\mu^2=-\frac{\phi}{2\alpha}=-\frac{\chi}{\phi}$ ($A,B$ are arbitrary).

Each of the cases from (a) to (e) above have an analogue in standard FLRW cosmology (without matter but with or without $\Lambda$). The only difference is, here the inflationary dynamics or a non-singular bounce could be supported even in the absence of a bare cosmological constant $\beta$, implying the possibility of a self-accelerating Universe. The last case (f) however, have no Einsteinian counterpart. This case also admits a nonsingular bounce for $A=B$.

\subsection*{Oscillatory solutions:}
These models are obtained only for $k=-1$ and $k=0$.

(a) $k=- 1$: $a(t)=A\cos{\nu t}+\sqrt{\frac{1}{\nu^2}-A^2} \sin{\nu t}$, $\nu^2=\frac{\phi}{2\alpha}\pm \sqrt{\left(\frac{\phi}{2\alpha}\right)^2-\frac{\chi}{2\alpha}}$. This represents a non-singular oscillatory Universe. 

(b) $k=-1$: $a(t)=\frac{1}{\nu}\sin{\nu t}$, $\nu^2=\frac{\phi}{2\alpha}\pm \sqrt{\left(\frac{\phi}{2\alpha}\right)^2-\frac{\chi}{2\alpha}}$. In each cycle, this oscillatory Universe exhibits a big bang (at $t=0$), reaches a maximum size (at $t=\frac{\pi}{2\nu}$) and recollapses to a singularity (at $t=\frac{\pi}{\nu}$).

(c) $k=0$: $a(t)=A\cos{\nu t}+B \sin{\nu t}$, 
$\nu^2=\frac{\phi}{2\alpha}=\frac{\chi}{\phi}$ ($A,B$ are arbitrary).

The cases (a) are (b) are similar to the FLRW solutions for $k=-1,\Lambda<0$, except the fact that here the solutions could exist even in the absence of a bare cosmological constant. Again, the last case (c) has no analogue in the vacuum Einsteinian case.

To conclude, all the $k=0$ or $k=+1$ solutions above are nonsingular. For any value of spatial curvature, the equation of motion admits a bounce followed by a smooth transition to a de-Sitter expansion (contraction) or vice versa.
%A de-Sitter and bouncing dynamics as well as their smooth superposition is allowed for all spatial curvatures as well and for a vanishing bare cosmological constant. 
In the absence of torsion, there is no power law Big-bang singularity.

\section{Static spherically symmetric solutions}
Here we shall be concerned with spherically symmetric static
%\footnote{The only propagating mode being a four-tensor, the assumption of staticity should be unnecessary since the Birkhoff's theorem is expected to hold.} 
solutions of the emergent theory, with the four-metric:
\begin{eqnarray*}
ds^2=-e^{\mu(r)}dt^2+e^{\lambda(r)} dr^2+r^2(d\theta^2+\sin^2 \theta d\phi^2)
\end{eqnarray*}
In terms of the radial functions $\mu(r),\lambda(r)$, eq.(\ref{eom}) becomes:
\begin{eqnarray}\label{eom3}
&&\phi\left[-\frac{1}{2}r^2\left(\mu''+\frac{\mu'}{2}(\mu'-\lambda')\right)-r\left(\mu'-\lambda'\right)+e^\lambda-1\right]\nonumber\\&&-~\alpha\left[\left(\mu''+\frac{\mu'}{2}(\mu'-\lambda')\right)(1-e^{-\lambda})
~+~\mu'\lambda'e^{-\lambda}\right]
~+~3\chi r^2 e^\lambda~=~0,
\end{eqnarray}
where we have used the following identities:
\begin{eqnarray*}
&&\epsilon^{abcd}\epsilon_{ijkl}e_a^i e_b^j\bar{R}_{cd}^{~kl}(\bar{w}(e))~=~
8\left[-\frac{1}{2}r^2\left(\mu''+\frac{\mu'}{2}(\mu'-\lambda')\right)-r\left(\mu'-\lambda'\right)+e^\lambda-1\right]e^{\frac{\mu-\lambda}{2}}\sin\theta,\nonumber\\
&&\epsilon^{abcd}\epsilon_{ijkl}\bar{R}_{ab}^{~ij}(\bar{w}(e))\bar{R}_{cd}^{~kl}(\bar{w}(e))~=~
-16\left[\left(\mu''+\frac{\mu'}{2}(\mu'-\lambda')\right)\left(1-e^{-\lambda}\right)+\mu'\lambda'e^{-\lambda}\right]e^{\frac{\mu-\lambda}{2}}\sin\theta.
\end{eqnarray*}
 This is the only equation relating two independent functions, reflecting an underdetermined system. Here, we shall focus on the question whether this admits Schwarzschild type solutions, i.e. satisfying $\mu(r)=-\lambda(r)$, and among them, if there could be black hole spacetimes.
 
 For this case, introducing the new variable $f(r)=e^{\mu(r)}-1=e^{-\lambda(r)}-1$, we may rewrite the field equation (\ref{eom3}) as:
 \begin{eqnarray}\label{feom}
\phi\left[\frac{r^2}{2}f''+2rf'+f\right]~-~ \alpha\left[ff''+f'^2\right] -3\chi r^2~=~0
 \end{eqnarray}
This has the following solution:
\begin{eqnarray}\label{fsol}
e^\mu(r)\equiv 1+f(r)=1+\frac{\phi}{2\alpha}r^2\pm \frac{1}{2}\sqrt{\left(\frac{\phi^2}{\alpha^2}-\frac{2\chi}{\alpha}\right)r^4+4C_1 r-4\frac{C_2}{\alpha}},
\end{eqnarray}
$C_1,C_2$ being integration constants. We note the formal similarity of the above with the Wiltshire class of metrics \cite{wiltshire} defined for $D\geq 5$ (the Boulware-Deser-Wheeler class \cite{wheeler} being a special case), which, however, were obtained in a different context and have no four-dimensional analogue.

  At the asymptotic limit, the above reduces to:
 \begin{eqnarray*}
 e^{\mu(r)}\rightarrow 1-\frac{\Lambda_{eff}r^2}{3}
 \pm \left[\frac{2M_{eff}}{r}- \frac{Q^2_{eff}}{r^2}\right] ~~(\mathrm{as} ~ r\rightarrow \infty)
 \end{eqnarray*}
 where we have defined the effective cosmological constant, Schwarzschild `mass' and Reissner-Nordstrom `charge', respectively as: $\Lambda_{eff}\equiv -\frac{3}{2}\left[\frac{\phi}{\alpha}\pm
  \sqrt{\frac{\phi^2}{\alpha^2}-\frac{2\chi}{\alpha}}\right],~2M_{eff}\equiv \left[\frac{C_1}{\sqrt{\frac{\phi^2}{\alpha^2}-\frac{2\chi}{\alpha}}}\right],~Q_{eff}^2\equiv \left[\frac{C_2}{\alpha \sqrt{\frac{\phi^2}{\alpha^2}-\frac{2\chi}{\alpha}}}\right]$, respectively. This solution is characterized by four parameters $(\frac{\alpha}{\phi},\Lambda_{eff},M_{eff},Q_{eff})$. The effective asymptotic mass is positive and the charge is real provided $C_1<0$ ($C_1>0$) and $\frac{C_2}{\alpha}<0$ ($\frac{C_2}{\alpha}>0$) for the `$+$' (`$-$') branch. As earlier, we assume $\frac{\phi^2}{\alpha^2}-\frac{2\chi}{\alpha}\geq 0$ so that the solutions are well-defined.
 
  The metric components are all finite as $r\rightarrow 0$. However, the emergent Ricci scalar $\bar{R}(\bar{w}(e))$ does diverge as $\frac{1}{r^2}$. To determine whether this curvature singularity lies at a finite proper time, let us consider the radial geodesic equation (timelike).  These are given by:
\begin{eqnarray*}
&&[1+f(r)]\left(\frac{dt}{d\tau}\right)^2 -\frac{1}{[1+f(r)]}\left(\frac{dr}{d\tau}\right)^2=1,~\frac{dt}{d\tau}=\frac{E}{1+f(r)},
\end{eqnarray*}
$E$ being a constant of motion. Hence, the proper time that elapses in reaching $r=0$ from any  finite distance $r=r_*$ is found to be:
\begin{eqnarray*}
\tau_*~=~|\int_{r_*}^{0}\frac{dr}{\sqrt{E^2-f(r)-1}}|
\end{eqnarray*}
In general, this cannot be integrated to a closed form. However, it is easy to show that there is at least a finite number of radial trajectories along which the singularity could be reached in finite proper time.
 Choosing the otherwise arbitrary integration constants as $C_1=0=C_2$, $\phi^2-2\alpha\chi=0$ for simplicity, we find:
 \begin{eqnarray*}
\tau_*~=~\frac{1}{\sqrt{\alpha}}\ln\left[\sqrt{\frac{\alpha}{E^2-1}}r_*+\sqrt{1+\frac{\alpha r_*^2}{E^2-1}}\right]
\end{eqnarray*} 
 This is finite.

 The solution (\ref{fsol}) represents a black hole spacetime, although different from the Schwarzschild geometry which is the unique spherically symmetric solution in Einsteinian case. In fact, the Schwarzschild spacetime is not an exact solution of the EOM (\ref{eom3}). 
The location of the horizon ($r_h$) is given by the following quartic equation:
 \begin{eqnarray}
 \chi r_h^4~+~2\phi r_h^2~-~2\alpha C_1 r_h~+~2(\alpha+ C_2)~=~0,
 \end{eqnarray}
 which admits more than one real roots in general (we avoid displaying their closed form expressions for brevity). For $\frac{\phi}{\chi}>0,~\frac{\alpha C_1}{\chi}<0,~\frac{\alpha+C_2}{\chi}>0$, all the terms above are monotonic in $r_h$, and hence admits one and only one horizon for any real root that exists. In order to understand whether a particular solution admits a single or double horizon, let us look at $\left(e^{\mu(r)}\right)'=\frac{\phi}{\alpha}r\pm \frac{2\left[C_1+\left(\frac{\phi^2}{\alpha^2}-\frac{2\chi}{\alpha}\right)r^3 \right]}{\sqrt{\left(\frac{\phi^2}{\alpha^2}-\frac{2\chi}{\alpha}\right)r^4+4C_1 r-\frac{4C_2}{\alpha}}}$. For the $`+'$ branch, this could take both positive and negative values at $r>0$ for both $\frac{\phi}{\alpha}>0$ or $\frac{\phi}{\alpha}<0$ and hence could admit a double horizon. For the $`-'$ branch, the double horizon could occur only if $\frac{\phi}{\alpha}>0$.
 
Let us also consider the case $C_1=0 =M_{eff}$ separately, as an example which does not come under the category mentioned in the last paragraph (no real root for $\frac{\phi}{\chi}>0,~\frac{\alpha+C_2}{\chi}>0$). This corresponds to a simpler expression for horizon radius:
 \begin{eqnarray*}
 r_h=\sqrt{-\frac{\phi}{\chi}\pm \sqrt{\frac{\phi^2}{\chi^2}-\frac{2(\alpha+C_2)}{\chi}}}
 \end{eqnarray*}
 For $\frac{\phi}{\chi}<0,~\frac{\alpha}{\chi}+\frac{C_2}{\chi}>0$, there are two horizons where the larger and smaller radius correspond to the $+$ and $-$ branch, respectively. For $\frac{\alpha}{\chi}+\frac{C_2}{\chi}<0$, there is exactly one horizon given by the $+$ branch.  The $\frac{\phi}{\chi}>0$, on the other hand, exhibits a horizon only if $\frac{\alpha}{\chi}+\frac{C_2}{\chi}<0$ and does not admit a double horizon.
 
We may look for more general static spherically symmetric solutions with $\mu(r)\neq -\lambda(r)$ parametrized  by a function $g(r)$ as: $e^{-\lambda}=g(r)e^\mu=g(r)(1+f(r))$ where $f(r)$ is given by (\ref{fsol}). By definition, $g(r)$ is a solution of the following first order equation obtained using eq.(\ref{eom3}) and (\ref{feom}):
\begin{eqnarray}
&&\phi\left[\frac{r^2 f' g'}{4}+r(1+f)g'-g(1-g)\right]~+~\alpha\left[\frac{gg'f'}{2}\left(1-3g(1+f)\right)+g(1-g)\left((1+f)f''+f^{'2}\right)\right]\nonumber\\
&&~-~3\chi r^2(1-g)~=~0
\end{eqnarray}
However, we have been unable to find any solution of the above other than $g=1$.

Note that the three-parameter black hole geometries obtained as a (singular) $D\rightarrow 4$ limit after regularization from the original $D\geq 5$ dimensional Einstein-Gauss-Bonnet field equations \cite{wheeler,konoplya,clifton}  emerge as a special case of the solutions (\ref{fsol}) here, with $C_2=0=Q_{eff}$. From the perspective of field equations, however, the earlier class of the configurations above should strictly be seen as either genuinely higher dimensional or as solutions of four dimensional scalar-tensor gravity \cite{lu,clifton} where the scalar mode does not decouple. In contrast, in our case the geometries (\ref{fsol}) are explicit solutions to the vacuum equations of motion (\ref{eom}) reflecting the effective Einstein-Gauss-Bonnet gravity, defined only by the four-metric and its (torsion-free) curvature.

\section{Higher Lovelock terms and uniqueness of emergent dynamics: }
In a generic Kaluza-Klein dimensional reduction of the Lovelock series, it is not only the Gauss-Bonnet, but rather, all higher lovelock terms could contribute to the four dimensional effective action in principle. Thus, there is no natural way to get a finite number of terms in the emergent equations of motion.
Here we explore if it is any different in our formulation based on extra dimensions of zero proper length, and  consider its generalization to higher dimensions ($D>5$) in this section.

 Beyond $D=5$, the next nontrivial case is $D=7$, when the cubic Lovelock term becomes dynamical. The corresponding Lagrangian density is given by:
\begin{eqnarray}\label{l7}
{\cal L}(\hat{e},\hat{w})&=&\epsilon^{\mu\nu\alpha\beta\gamma\delta\lambda} \epsilon_{IJKLMNP}~[\sigma \hat{R}_{\mu\nu}^{~~IJ}(\hat{w})\hat{R}_{\alpha\beta}^{~~KL}(\hat{w})  \hat{R}_{\gamma\delta}^{~~MN}(\hat{w})~+~\frac{\alpha}{6}~\hat{R}_{\mu\nu}^{~~IJ}(\hat{w})\hat{R}_{\alpha\beta}^{~~KL}(\hat{w})\hat{e}_{\gamma}^{M}  \hat{e}_{\delta}^{N}\nonumber\\
&+&~\frac{\zeta}{5}~\hat{R}_{\mu\nu}^{~~IJ}(\hat{w})\hat{e}_{\alpha}^{K}\hat{e}_{\beta}^{L}\hat{e}_{\gamma}^{M} \hat{e}_{\delta}^{N}+\frac{\beta}{7}~\hat{e}_{\mu}^{I}\hat{e}_{\nu}^{J}\hat{e}_{\alpha}^{K}\hat{e}_{\beta}^{L}\hat{e}_{\gamma}^{M}\hat{e}_{\delta}^{N}]~\hat{e}_\lambda^P,
\end{eqnarray} 
The connection and vielbein equations read, respectively:
 \begin{eqnarray}
&&\epsilon^{\mu\nu\alpha\beta\gamma\delta\lambda} \epsilon_{IJKLMNP}~\left[3\sigma
\hat{R}_{\mu\nu}^{~~IJ}(\hat{w})\hat{R}_{\alpha\beta}^{~~KL}(\hat{w})~+~\alpha\hat{e}_{\mu}^{I}\hat{e}_{\nu}^{J}\hat{R}_{\alpha\beta}^{~~KL}(\hat{w})
~+~
\zeta\hat{e}_{\mu}^{I} \hat{e}_{\nu}^{J}\hat{e}_{\alpha}^{K}\hat{e}_{\beta}^{L}\right]\left(\hat{D}_{\gamma}(\hat{w})\hat{e}_\delta^N\right)=0,\nonumber\\
\label{eom4}\\
%%%%%%%
%%%%%%%
&&\epsilon^{\mu\nu\alpha\beta\gamma\delta\lambda} \epsilon_{IJKLMNP}~[\sigma
\hat{R}_{\mu\nu}^{~~IJ}(\hat{w})\hat{R}_{\alpha\beta}^{~~KL}(\hat{w})\hat{R}_{\gamma\delta}^{~~MN}(\hat{w})~+~\frac{\alpha}{2}\hat{e}_{\mu}^{I}\hat{e}_{\nu}^{J}\hat{R}_{\alpha\beta}^{~~KL}(\hat{w})\hat{R}_{\gamma\delta}^{~~MN}(\hat{w})\nonumber\\
&&+~ \zeta\hat{e}_{\mu}^{I} \hat{e}_{\nu}^{J}\hat{e}_{\alpha}^{K}\hat{e}_{\beta}^{L}\hat{R}_{\gamma\delta}^{~~MN}(\hat{w})
~+~
\beta\hat{e}_{\mu}^{I} \hat{e}_{\nu}^{J}\hat{e}_{\alpha}^{K}\hat{e}_{\beta}^{L}\hat{e}_{\gamma}^{M} \hat{e}_{\delta}^{N}]=0
\label{eom5}
\end{eqnarray}
The connection equations of motion (\ref{eom4}) are solved by: 
 \begin{eqnarray}\label{torsion}
&&\epsilon^{\mu\nu\alpha\beta\gamma\delta\lambda} \epsilon_{IJKLMNP}\left[3\sigma
\hat{R}_{\mu\nu}^{~~IJ}(\hat{w})\hat{R}_{\alpha\beta}^{~~KL}(\hat{w})~+~\alpha\hat{e}_{\mu}^{I}\hat{e}_{\nu}^{J}\hat{R}_{\alpha\beta}^{~~KL}(\hat{w})
~+~
\zeta\hat{e}_{\mu}^{I} \hat{e}_{\nu}^{J}\hat{e}_{\alpha}^{K}\hat{e}_{\beta}^{L}\right]=0 ~~{\mathrm or,~}\nonumber\\
&&\hat{D}_{[\alpha}(\hat{w})e_{\beta]}^{I}=0.
\end{eqnarray}
It is straightforward to verify that the first case above, involving a quadratic curvature contribution, reproduces the effective equation of motion (\ref{eom}) after a redefinition of couplings. The second case,
implying vanishing torsion in the $D$-dimensional spacetime, needs a more detailed analysis as presented below. 

Since the full spacetime now has three dimensions of vanishing proper length associated with the three zero eigenvalues of the seven dimensional vielbein, let us introduce a more general notation as follows. The spacetime and internal indices respectively are decomposed  as: $\mu\equiv(a,\bar{a})$, $I\equiv (i,\bar{i})$ where $a,i$ are the four dimensional indices and $\bar{a}\equiv (v_1,v_2,v_3),~\bar{i}\equiv (4,5,6)$ are the extra dimensional ones. $\hat{e}_a^i\equiv e_a^i$ denotes the emergent tetrad as before (with emergent inverse tetrads denoted as $e^a_i$), whereas $\hat{e}_{\bar{a}}^{\bar{i}}=0=\hat{e}_a^{\bar{i}}=\hat{e}_{\bar{a}}^i$.

With this, the eqs.(\ref{torsion}) are solved for the super-connection field components as:
\begin{eqnarray}
\hat{D}_{[a}(\hat{w})e_{b]}^{i}&=&0\Rightarrow K_a^{~ij}\equiv \hat{w}_a^{~ij}-\bar{w}_a^{~ij}(e)=0,\nonumber\\
\hat{D}_{[a}(\hat{w})e_{b]}^{\bar{i}}&=& 0\Rightarrow \hat{w}_a^{~\bar{i}i}=\overset{\bar{(i)}}{{Q}^{ik}}e_{ak}~[\overset{\bar{(i)}}{{Q}^{ik}}=\overset{\bar{(i)}}{{Q}^{ki}}],\nonumber\\
%%%%%%
\hat{D}_{[\bar{a}}(\hat{w})e_{b]}^{\bar{i}}&=& 0\Rightarrow \hat{w}_{\bar{a}}^{~\bar{i}i}=0,
\nonumber\\
\hat{D}_{[\bar{a}}(\hat{w})e_{b]}^{i}&=& 0\Rightarrow \hat{w}_{\bar{a}}^{~ij}=-e^a_j\del_{\bar{a}} e_a^i.
\end{eqnarray}
 The remaining equations $\hat{D}_{[\bar{a}}(\hat{w})e_{\bar{b}]}^{I}=0$ are identically satisfied. Note that the last equation above imply that $\hat{w}_{\bar{a}}^{~ij}$ is a pure gauge and the tetrad determinant (and hence the emergent metric) is independent of the extra dimensional coordinates, a result analogous to the  five dimensional case analyzed in the previous sections. While these could be gauged away: $\hat{w}_{\bar{a}}^{~ij}=0$,  the components $\hat{w}_{\bar{a}}^{~\bar{i} \bar{j}},~\hat{w}_{a}^{~\bar{i} \bar{j}}$ are left arbitrary.

 From the above solutions for connection fields, we find that the following field-strength components are trivial:
 \begin{eqnarray}
\hat{R}_{\bar{a}a}^{~~ij}=0,~ \hat{R}_{\bar{a}\bar{b}}^{~~ij}=0,
~\hat{R}_{\bar{a}\bar{b}}^{~~\bar{i}i}=0.
 \end{eqnarray}
 Further, let us consider the identity:
 \begin{eqnarray}
 \hat{R}_{a\bar{a}}^{~~i\bar{i}}=\hat{w}_{a}^{~i\bar{k}}\hat{w}_{\bar{a}}^{~\bar{i}\bar{k}}
 \end{eqnarray}
This, along with the antisymmetry of $\hat{R}_{a\bar{a}}^{~~i\bar{i}}=-\hat{R}_{a\bar{a}}^{~~\bar{i}i}$ forces it to vanish: $\hat{R}_{a\bar{a}}^{~~i\bar{i}}=0$. This condition admits three possible solutions: \\
a) $\hat{w}_{\bar{a}}^{~\bar{i}\bar{k}}=0$;\\
b) $\hat{w}_{a}^{~i\bar{k}}=0=\overset{\bar{(i)}}{{Q}^{ik}}$;\\
c) At least one component of the connection fields $\hat{w}_{a}^{~i\bar{k}}$ is nonvanishing. For instance, we may choose:
$\hat{w}_{\bar{a}}^{~\bar{i}4}=0$, $\hat{w}_{a}^{~i\bar{k}}=Q_a^i \delta_4^{\bar{k}}$ ($Q_a^i\equiv Q^{ij}e_{aj},~Q^{ij}=Q^{ji}$) while leaving $\hat{w}_{\bar{a}}^{~\bar{i}5},~\hat{w}_{\bar{a}}^{~\bar{i}6}$ arbitrary. \\
We consider all three cases next. 
 
 Using the expressions for the field-strength components, all the components of the vielbein equations of motion eq.(\ref{eom5}) are identically satisfied except $(\lambda,P)=(\bar{c},p)$ and $(\lambda,P)=(\bar{c},\bar{p})$, leading to, respectively:
 \begin{eqnarray}\label{eom8}
\epsilon^{abcd\bar{a}\bar{b}\bar{c}} \epsilon_{ijkp\bar{i}\bar{j}\bar{k}}~\left[\sigma
\hat{R}_{ab}^{~~ij}+\alpha e_a^i e_b^j\right]\hat{R}_{cd}^{~~k\bar{i}}\hat{R}_{\bar{a}\bar{b}}^{~~\bar{j}\bar{k}}=0,\nonumber\\
\epsilon^{abcd\bar{a}\bar{b}\bar{c}} \epsilon_{ijkl\bar{i}\bar{j}\bar{p}}~\left[\sigma
\hat{R}_{ab}^{~~ij}\hat{R}_{cd}^{~~kl}+\frac{\alpha}{2} e_a^i e_b^j\hat{R}_{cd}^{~~kl}+\zeta e_a^i e_b^j e_c^k e_d^l\right]\hat{R}_{\bar{a}\bar{b}}^{~~\bar{i}\bar{j}}=0
\end{eqnarray}  

Let consider case (c) first, implying $\hat{R}_{\bar{a}\bar{b}}^{~~\bar{i}4}=0,~\hat{R}_{ab}^{~~i4}=\hat{D}_{[a}(\hat{w})M_{b]}^i$ and leaving $\hat{R}_{ab}^{~~i5},~ \hat{R}_{ab}^{~~i6},~\hat{R}_{\bar{a}\bar{b}}^{~~56}$ arbitrary. Using these, we obtain from eq.(\ref{eom8}):
\begin{eqnarray}\label{eom9}
\epsilon^{abcd} \epsilon_{ijkl}~\left[\sigma
\hat{R}_{ab}^{~~ij}+\alpha e_a^i e_b^j\right]\bar{D}_{c}(\bar{w})Q_{d}^i=0,\nonumber\\
\epsilon^{abcd} \epsilon_{ijkl}~\left[\sigma
\hat{R}_{ab}^{~~ij}\hat{R}_{cd}^{~~kl}+\frac{\alpha}{2} e_a^i e_b^j\hat{R}_{cd}^{~~kl}+\zeta e_a^i e_b^j e_c^k e_d^l\right]=0.
\end{eqnarray}  
 where we have defined $\epsilon^{abcd v_1 v_2 v_3}\equiv \epsilon^{abcd}$ and $\epsilon_{ijkl456}\equiv \epsilon_{ijkl}$. The first among the set above is solved by:
 \begin{eqnarray}\label{eom10}
\sigma\hat{R}_{ab}^{~~ij}+\frac{\alpha}{2} e_a^{[i} e_b^{j]}&=&0 ~~\mathrm{or,}\\
 Q_{a}^i&=&\lambda e_a^i.\label{eom11}
 \end{eqnarray}
 The first solution (\ref{eom10}) represents maximally symmetric spacetimes. Using this in the second equation in (\ref{eom9}), we obtain: $\alpha^2+2\sigma\zeta=0$. This case is thus equivalent to the constant curvature solutions already discussed in section-III, as reflected by the exact correspondence with equations (\ref{constk}) and (\ref{coupling}).
 The other solution (\ref{eom11}), when inserted back into the second equation in (\ref{eom9}), precisely reproduces the equation of motion (\ref{eom}) in section-IV after a redefinition of the couplings. Hence, this class corresponds to the vanishing torsion solutions for the five dimensional case analyzed in section-IV.  

It is now easy to verify that the case (b), which implies $\hat{R}_{ab}^{~~k\bar{i}}=0$, represents the limit of the above equation of motion (\ref{eom9}) where $Q_a^i=0$.  
Case (a) on the other hand implies $ \hat{R}_{\bar{a}\bar{b}}^{~~\bar{i}\bar{j}}=0$, satisfying both the equations in (\ref{eom8}) identically. This leads to a dynamically trivial emergent theory and may simply be discarded.  

From these results, we conclude that a $D=7$ theory with three additional dimensions of vanishing proper length and containing a cubic Lovelock term in the action reduces to the five dimensional case. In other words, only the quadratic curvature correction shows up in the four dimensional effective theory. Note, however, that the couplings are now shifted in the sense that the coefficient of the quadratic curvature contribution to the field equation (\ref{eom9}) corresponds to the cubic term in the $D=7$ Lagrangian density (\ref{l7}). Cubic curvature corrections could, for instance, emerge in a six dimensional effective theory with a $D=7+2n$ degenerate vielbein with $1+2n$ null eigenvalues, a case that falls outside the context of the present work. 

 A generalization of this result to any $D>7$ is straightforward, and does not change this conclusion. To emphasize, the pattern that a $D+2$ dimensional theory ($D\geq 5$) simply reproduces the $D$ dimensional equations of motion keeps repeating itself. Thus, one need not  consider any higher order Lovelock term other than Gauss-Bonnet, so long as the number of dimensions having a nonvanishing metrical length remains exactly four.  %To emphasize, the Gauss-Bonnet term emerges as the only nonlinear Lovelock term relevant for the four dimensional effective theory in our formulation, 
%This implies that the effective equation exhibits only a quadratic nonlinearity in curvature, as opposed to an infinity of higher ones.

%This result signifies an enormous simplication, as well as an important contrast to the classes of four dimensional effective theories based on the Kaluza-Klein reduction of higher dimensional Lovelock terms or four dimensional higher curvature theories in general. 
%In these cases, in an arbitrary number of dimensions, the full series should contribute in principle, implying an infinite number of possible correction terms in an effective equation of motion in general.

\section{Conclusions}
Inspired by the idea of extra dimension of zero proper length, introduced recently \cite{sengupta} within a first order formulation, we have obtained a unique quadratic gravity theory carrying the dynamical imprint of Gauss-Bonnet density in four (effective) dimensions.
 This formalism is inequivalent to a Kaluza-Klein reduction, as the extra dimension is not associated with a physical evolution and does not lead to any infinite tower of discrete eigenmodes of higher energy. The effective four dimensional equations of motion are purely geometric, and have no higher than second order derivatives of the four-metric. 

Notably, the field equations defining the 4D quadratic curvature gravity theory
remains unaffected by the inclusion of any arbitrary number of higher Lovelock terms in $D=5+2n$ dimensions in general. In other words, increasing the number of `dark' extra dimensions (of vanishing proper length) has no effect on the emergent theory, as long as the effective spacetime is defined by an invertible tetrad. 
 This signifies an enormous simplication, particularly when contrasted with approaches based on ghost-free higher curvature theories \cite{biswas} or a Kaluza-Klein dimensional reduction where the effective equations of motion could exhibit an infinity of nonlinear terms in principle. 

%T In general, there is a natural connect between this theory to one with two dynamical metrics. However, this observation demands a detailed analysis, and could be taken up elsewhere.  Here, we are concerned with the case where there is no propagating degree of freedom other than a single emergent metric. 

The five dimensional equations are solved by spacetimes with maximal symmetry and vanishing torsion. Both cases have been worked out in detail. To emphasize, this formulation does not require any singular rescaling of the couplings or any regularization of divergent terms. This is in contrast with a recent proposal by Glavan and Lin \cite{glavan}, as well as with others based on a Kaluza-Klein reduction or conformal scaling \cite{lu,clifton}, which reproduce variants of scalar-tensor gravity in general. 

Apart from the general motive of our construction, the details reveal some special features in the context of homogeneous and isotropic (FLRW) cosmology. For instance, even in absence of a bare cosmological constant and any spatial curvature, the cosmological field equations could exhibit nonsingular bounce or inflationary behaviour. This has no analogue in standard (Einsteinian) FRW cosmology. In general, the dynamics admits smooth transitions between a bounce and either an inflationary universe or one undergoing a smooth contraction. That these features could be obtained in vacuum theories with Gauss-Bonnet induced nonlinearities only might turn out to be a more economical alternative for bouncing cosmology models, where it is imperative to choose an appropriate  scalar potential (typically exponential for a slow contraction \cite{steinhardt}) out of many other possibilities.

Further, we have shown that this formulation admits a family of static, spherically symmetric black hole solutions in vacuum which are more general than the Schwarzschild-de Sitter class characterizing Einstein gravity. The four parameters defining each such solution correspond to (effective) Schwarzschild mass, charge, cosmological constant and the Gauss-Bonnet coupling. In the asymptotic limit, the Schwarzschild geometry emerges alongwith a higher order effect reflected by a geometric `charge'.

An intriguing question is, whether the extra dimensional formalism \cite{sengupta} adopted here could admit an emergent electromagnetic field strength, as does the Kaluza-Klein proposal, inspite of the fundamental differences between these two formulations. However, the finer details do reveal that a similar description could also be obtained with vanishing metrical dimensions, provided the higher dimensional connection fields are interpreted appropriately \cite{sengupt}. In fact, the natural emergence of the purely geometric ‘charge’ $Q_{eff}$ does suggest
so. Elaborations on these aspects would be reserved for future work.

We cannot claim our analysis to be complete. While we have answered the question as to why it should be sufficient to consider only one invisible extra dimension, we have not touched upon the more fundamental one as to why should the emergent spacetime exhibit exactly four dimensions with a noninvertible tetrad. We have left the case of nonvanishing torsion \cite{mardones}
 for future work. Nor have we included any matter coupling. In the cosmological context, torsional degrees of freedom could as well introduce effects akin to matter, leaving open the possibility of a richer cosmological dynamics (potentially nonsingular). A detailed Hamiltonian analysis of this theory should also be taken up elsewhere.

To conclude, our studies here could be relevant from the general perspective of possible topology change of lower (four) dimensional spacetimes mediated by degenerate metric solutions of first order gravity.
% in classical gravity theory or otherwise. 
Possibly, the hidden (`dark') dimensions could have played a role during the `beginning' of the Universe, if there has been one. In this regard, it seems essential to analyze the contribution of such topology changing processes to the quantum gravitational dynamics. 
In fact, taking up such questions would be in line with insights gained from canonical quantization approaches to gravity, which seem to suggest that the notion of a smooth spacetime defined by an invertible tetrad is unlikely to survive in the quantum theory. 

% To conclude, some of the insights gained from this work should serve as sufficient motivation for future investigations along these lines.

%\begin{acknowledgments}
\acknowledgments 
The author wishes to acknowledge the support (in part) of the SERB, DST, Govt. of India., through the MATRICS project grant MTR/2021/000008 (approved recently).
%\end{acknowledgments}


\begin{thebibliography}{99}

\bibitem{sengupta} S. Sengupta, Gravity theory with a dark extra dimension,
Phys. Rev. D 101, 104040 (2020) [arXiv:1908.04830 [gr-qc]].

\bibitem{lovelock} D. Lovelock, The Einstein tensor and its generalizations, J. Math. Phys. 12 (1971) 498;\\
D. Lovelock, The four-dimensionality of space and the
Einstein tensor, J. Math. Phys. 13, (1972) 874.
%;\\
\bibitem{lanc} C. Lanczos, A Remarkable property of the Riemann-Christoffel tensor in four dimensions, Annals Math. 39 (1938) 842.

\bibitem{kaluza} T. Kaluza, Sitzungsber. Preuss. Akad. Wiss. Phys.
Math. Klasse 996 (1921).

\bibitem{klein} O. Klein, Z. F. Physik 37 (1926) 895;\\
O. Klein, Nature 118 (1926) 516.

\bibitem{hoissen} J. Madore, Kaluza-Klein theory with the Lanczos lagrangian, Phys.Lett.A 110 (1985) 289;\\
F. Muller-Hoissen, Dimensionally continued Euler forms: Kaluza-Klein cosmology and dimensional reduction, Class. Quantum Grav. 3 (1986) 665;\\
F. Muller-Hoissen, Spontaneous Compactification With Quadratic and Cubic Curvature Terms, Phys.Lett.B 163 (1985) 106.

\bibitem{deruelle} N. Deruelle and L. Fariña-Busto, Lovelock gravitational field equations in cosmology, Phys. Rev. D 41, 3696 (1990).

\bibitem{felisola} O. Castillo-Felisola et al, Kaluza-Klein cosmology from five-dimensional Lovelock-Cartan theory, Phys. Rev. D 94, 124020 (2016) [arXiv:1609.09045 [gr-qc]].

\bibitem{wheeler} D. G. Boulware and S. Deser, String Generated Gravity
Models, Phys. Rev. Lett. 55, 2656 (1985);\\
J. T. Wheeler, Symmetric solutions to the Gauss-Bonnet extended Einstein equations, Nucl. Phys. B 268, 737 (1986); Symmetric solutions to the maximally Gauss-Bonnet
extended Einstein equations, Nucl. Phys. B 273, 732 (1986).


\bibitem{glavan} D. Glavan and C. Lin, Einstein-Gauss-Bonnet Gravity
in Four-Dimensional Spacetime, Phys. Rev. Lett. 124 [arXiv:1905.03601 [gr-qc]].

\bibitem{tekin} M. Gurses, T. Cagri Sisman, and B. Tekin, Comment on ``Einstein-Gauss-Bonnet Gravity in 4-Dimensional Space-Time'' , Phys. Rev. Lett. 125, 149001 (2020) [arXiv:2009.13508 [gr-qc]];\\
 M. Gurses, T. Cagri Sisman, and B. Tekin, Is there a novel Einstein–Gauss–Bonnet theory in four dimensions?, Eur. Phys. J. C 80 (2020) 7, 647 [arXiv:2004.03390 [gr-qc]].
 
\bibitem{arrechea} J. Arrechea, A. Delhom, and A. Jiménez-Cano, 
Comment on ``Einstein-Gauss-Bonnet Gravity in Four-Dimensional Spacetime'',
Phys. Rev. Lett. 125, 149002 (2020) [arXiv:2009.10715 [gr-qc]];\\
J. Arrechea, A. Delhom, and A. Jiménez-Cano, Inconsistencies in four-dimensional Einstein-Gauss-Bonnet gravity, arXiv:2004.12998 [gr-qc].

\bibitem{aoki} K. Aoki, M. A. Gorji, S. Mukohyama, A consistent theory of $D\to 4$ Einstein-Gauss-Bonnet gravity, Phys. Lett. B 810 (2020) 135843 [arXiv:2005.03859 [gr-qc]];\\
Wen-Yuan Ai, A note on the novel 4D Einstein–Gauss–Bonnet gravity, Commun.Theor.Phys. 72 (2020) 9, 095402 [arXiv:2004.02858 [gr-qc]].

\bibitem{horndeski} G. W. Horndeski, “Second-order scalar-tensor field equations in a four-dimensional space,”
Int. J. Theor. Phys. 10 (1974) 363.

\bibitem{lu} H. Lu and Y. Pang, Horndeski gravity as $d\to 4$ limit of
Gauss-Bonnet, Phys. Lett .B 809 (2020) 135717 [arXiv:2003.11552 [gr-qc]];\\
T. Kobayashi, Effective scalar-tensor description of regularized Lovelock gravity in four dimensions, JCAP 07 (2020) 013 [arXiv:2003.12771 [gr-qc]];\\
J. Bonifacio, K. Hinterbichler, and L. A. Johnson, Amplitudes and 4D Gauss-Bonnet Theory, Phys. Rev. D 102, 024029 (2020) [arXiv:2004.10716 [hep-th]].


\bibitem{mann}  R. B. Mann and S. F. Ross, The $D \to 2$ limit of general relativity, Class. Quant. Grav. 10 (1993) 1405 [arXiv:gr-qc/9208004].

\bibitem{clifton} Pedro G. S. Fernandes, Pedro Carrilho, Timothy Clifton, David J. Mulryne, Derivation of Regularized Field Equations for the Einstein-Gauss-Bonnet Theory in Four Dimensions, Phys. Rev. D 102, 024025 (2020) [arXiv:2004.08362 [gr-qc]];\\
R. A. Hennigar, D. Kubiznak, R. B. Mann, C. Pollack, On Taking the $D\to4$ limit of Gauss-Bonnet Gravity: Theory and Solutions, J. High Energ. Phys. 2020, 27 (2020) [arXiv:2004.09472 [gr-qc]];\\
D. A. Easson, T. Manton and Andrew Svesko, 
    $D\to4$ Einstein-Gauss-Bonnet Gravity and Beyond, JCAP 10 (2020) 026 [arXiv:2005.12292 [hep-th]].
    
\bibitem{horowitz} G. T. Horowitz, Topology change in classical and quantum gravity, Class. Quant. Grav. 8 (1991) 587;\\
I. Bengtsson, Some observations on degenerate metrics, Gen. Relat. Gravit. 25, 101 (1993). 

\bibitem{tseytlin} A. A. Tseytlin, On the first-order formalism in quantum gravity, J. Phys. A: Math. Gen. 15 (1982) L105.
%\bibitem{horowitz} G. T. Horowitz, Class. Quant. Grav. 8 (1991) 587-602.

\bibitem{kaul} R. K. Kaul and S. Sengupta, Degenerate spacetimes in first order gravity, Phys. Rev. D 93, 084026
(2016);\\
R. K. Kaul and S. Sengupta, New solutions in pure gravity with degenerate tetrads, Phys. Rev. D 94, 104047
(2016).
%\bibitem{konoplya} R. A. Konoplya and A. Zhidenko, Black holes in the four-dimensional Einstein-Lovelock gravity, Phys. Rev. D 101, 084038 (2020) [arXiv:2003.07788 [gr-qc]]

\bibitem{biswas} T. Biswas, E. Gerwick, T. Koivisto and A. Mazumdar, Towards singularity and ghost free theories of gravity, Phys. Rev. Lett. 108 (2012) 031101 [arXiv:1110.5249 [gr-qc]].

\bibitem{steinhardt} A. Ijjas and P.J. Steinhardt, Bouncing Cosmology made simple, Class.Quant.Grav. 35 (2018) 13, 135004 [arXiv:1803.01961 [astro-ph.CO]].

\bibitem{biswas1} T. Biswas, A. Mazumdar, W. Siegel, Bouncing universes in string-inspired gravity, JCAP 03 (2006) 009 [arXiv:hep-th/0508194].


%\bibitem{albert} A. Einstein and P. Bergmann, Annals Math. 39 (1938) 683-701.


%\bibitem{dvali} N. Arkani-Hamed, S. Dimopoulos and G. Dvali, Phys.
%Lett. B 429 (1998) 263;\\
%I. Antoniadis, N. Arkani-Hamed, S. Dimopoulos and G. Dvali, Phys. Lett. B 436 (1998) 257.

%\bibitem{rs} L. Randall and R. Sundrum, Phys. Rev. Lett. 83 (1999)
%3370-3373.

%\bibitem{rs1} L. Randall and R. Sundrum, Phys. Rev. Lett. 83 (1999) 4690-4693.

%\bibitem{ant} I. Antoniadis, Phys. Lett. B 246 (1990) 377-384.

\bibitem{wiltshire} D. L. Wiltshire, Spherically Symmetric Solutions of Einstein-Maxwell Theory With a {Gauss-Bonnet} Term, Phys. Lett. B 169 (1986) 36.

\bibitem{konoplya} R. A. Konoplya and A. F. Zinhailo, Quasinormal modes, stability and shadows of a black hole in the 4D Einstein-Gauss-Bonnet gravity,  Eur. Phys. J. C 80 (2020) 11, 1049 [arXiv:2003.01188 [gr-qc]];\\
%
R. Roy and S. Chakrabarti, A study on black hole shadows in asymptotically de Sitter spacetimes, Phys. Rev. D 102 (2020) 2, 024059 [arXiv:2003.14107 [gr-qc]];\\
%
N. Dadhich, On causal structure of 4D-Einstein-Gauss-Bonnet black hole, Eur.
Phys. J. C 80 (2020) 9, 832 [arXiv:2005.05757 [gr-qc]].


\bibitem{sengupt} S. Gera and S. Sengupta, Finite model of an electric charge, Phys. Rev. D 104, 044057 (2021);\\ S. Gera and S. Sengupta, Magnetic monopole as a spacetime defect, Phys. Rev. D 104, 044038 (2021).

\bibitem{mardones} A. Mardones and J. Zanelli, Lovelock-Cartan theory of gravity, Class. Quant. Grav. 8 (1991) 1545.

%\bibitem{san} S. Sengupta, Phys. Rev. D 97, 124038 (2018);\\
%S. Gera and S. Sengupta. Phys. Rev. D 99, 124038 (2019).



\end{thebibliography}
\end{document}